\documentclass[
  aps,
  pre, 
  reprint, 
]{revtex4-2}

\usepackage[version=3]{mhchem}
\usepackage{mathtools,bm}
\usepackage{hyperref}
\usepackage{xcolor}

\usepackage{graphicx}
\usepackage{dcolumn}
\usepackage{bm}

\usepackage{amsmath, bm, amssymb}
\newcommand{\bc}{\bm{c}}

\newcommand{\bx}{\bm{x}}
\newcommand{\bX}{\bm{X}}
\newcommand{\bu}{\bm{u}}
\newcommand{\bP}{\bm{P}}
\newcommand{\bR}{\bm{R}}
\newcommand{\bS}{\bm{S}}

\newcommand{\bI}{\bm{I}}

\newcommand{\bnabla}{\bm{\nabla}}

\newcommand{\bOmega}{\bm{\Omega}}
\newcommand{\od}{\text{d}}

\begin{document}

\title{Self-organization of active colloids mediated by chemical interactions}

\author{Zhiwei Peng}
\thanks{Present Address: Department of Chemical and Materials Engineering, University of Alberta, Edmonton, AB, T6G 1H9, Canada. Email: zhiwei.peng@ualberta.ca}
\author{Raymond Kapral}%
 \email{r.kapral@utoronto.ca}
\affiliation{%
Chemical Physics Theory Group, Department of Chemistry, University of Toronto,
Toronto, ON, M5S 3H6, Canada
}%

\date{\today}
\begin{abstract}
  Self-propelled colloidal particles exhibit rich non-equilibrium phenomena and have promising applications in fields such as drug delivery and self-assembled active materials. Previous experimental and theoretical studies have shown that chemically active colloids that consume or produce a chemical can self-organize into clusters with diverse characteristics depending on the effective phoretic interactions. In this paper, we investigate self-organization in systems with multiple chemical species that undergo a network of reactions and multiple colloidal species that participate in different reactions.  Active colloids propelled by complex chemical reactions with potentially nonlinear kinetics can be realized using enzymatic reactions that occur on the surface of enzyme-coated particles.  To demonstrate how the self-organizing behavior depends on the chemical reactions active colloids catalyze and their chemical environment, we consider first a single type of colloid undergoing a simple catalytic reaction, and compare this often-studied case with self-organization in binary mixtures of colloids with sequential reactions, and binary mixtures with nonlinear autocatalytic reactions. Our results show that in general active colloids at low particle densities can form localized clusters in the presence of bulk chemical reactions and phoretic attractions. The characteristics of the clusters, however, depend on the reaction kinetics in the bulk and on the particles and phoretic coefficients. With one or two chemical species that only undergo surface reactions, the space for possible self-organizations are limited. By considering the additional system parameters that enter the chemical reaction network involving reactions on the colloids and in the fluid, the design space of colloidal self-organization can be enlarged, leading to a variety of non-equilibrium structures. 
\end{abstract}

\maketitle


\section{Introduction}
Self-organizing dynamics is often observed in active matter systems across a wide range of length scales. Examples of self-organization in active matter include aster formation of molecular motors and microtubules \cite{ndlec1997self,surrey2001physical,lee2001macroscopic,ross2019controlling}, swarming of bacteria \cite{kearns2010field,ariel2015swarming,be2019statistical}, motility-induced phase separation of self-propelled colloids \cite{Fily_Marchetti_12,Redner_2013,Cates2015,Takatori_2015,Solon_2018,Hermann_2019,Omar_2023,Zhao23}, and flocking behavior in birds and fish \cite{toner1998flocks,cavagna2014bird,bialek2012statistical,weihs1973hydromechanics}. Through internal interactions, the constitutive agents or particles of these systems are able to produce emergent patterns, structures, or order. Among such interactions, chemical interaction is often employed by both biological and synthetic active particles. Motile microorganisms and cells can perform chemotaxis in response to chemical concentration gradients \cite{berg1975chemotaxis,adler1975chemotaxis,keller1971model}. In addition to responding to a concentration gradient imposed in the environment, they may also produce or consume the same chemical species, thus modifying the chemical concentration in the environment. Synthetic diffusiophoretic colloids can move and reorient in response to chemical gradients, mimicking the chemotactic behavior of microorganisms. 

Due to their ability to self-organize into dynamic structures, active colloids are useful building blocks for self-assembled materials \cite{Mallory2018}. Previous experimental and theoretical studies have shown that phoretic active colloids can form clusters even at low particle densities \cite{PRL2012,TK12,palacci2013living,Pohl_Stark_2014,pohl2015self,HSGKJCP:2019,Varma18,Giovanni19,meredith2020predator}. Consider a self-diffusiophoretic particle that consumes a chemical in a reaction on its surface and, in the process, creates a nonuniform chemical concentration field. The chemical concentration gradients enable diffusiophoresis in which particles may effectively attract or repel each other depending on the details of the chemical interaction. If the particles experience a diffusiophoretic force towards regions of lower concentration, they tend to attract each other since they are consuming the chemical and act as chemical sinks. Examples of the complex non-equilibrium dynamics systems of this type exhibit, that include cluster and aster formation, as well as plasma-like oscillations in response to perturbations, are given in \citet{Saha2014} who considered the chemotactic response self-diffusiophoretic colloids to two chemical species $S$ and $P$ in the reaction $S \to P$. Chemotactic aggregation or collapse has been observed in biological active matter \cite{KELLER1970399,brenner1998physical,mittal2003motility,Peruani12}. 
When the translational and rotational phoretic motion give competing attractive and repulsive interactions, dynamic clustering states can be observed. Other variants of this model focusing on rotational phoretic interaction also give rise to pattern formation and clustering \cite{Liebchen2015,Liebchen2017}; in this case, the active particles tend to align with or against the gradient of the chemical field in addition to their self-propulsion. Continuum theories have been developed that produce clustering and pattern formation dynamics similar to the particle-based models \cite{Pohl_Stark_2014,Saha2014,Liebchen2015,Liebchen2017,GK-Research2020}. 

So far we have outlined the dynamics of a single type of colloid interacting via a single chemical species. That is, all colloids in the system respond to the chemical field and produce or consume the chemical in the same fashion. A natural extension of these previous studies is to consider the dynamics of mixtures of colloids that interact with the single chemical species differently. In a binary mixture of such colloids, the effective action-reaction symmetry can be broken, which leads to non-reciprocal interactions between the two types of colloids. Indeed, previous work has shown that such mixtures exhibit aggregation or phase-separated states with distinct densities depending on the interaction parameters \cite{Agudo2019}. 

In this paper, we generalize previous studies by considering a model for active colloidal mixtures that interact chemically with multiple chemical species that participate in a network of chemical reactions. Instead of producing or consuming a single chemical, the colloids may participate in multiple chemical reactions, thus potentially producing or consuming multiple chemical species. By allowing the particles and fluid to participate in multiple coupled reactions, the design space for colloidal self-organization and self-assembly is greatly enlarged. Active colloids propelled by complex chemical reactions with potentially nonlinear kinetics can be realized using enzymatic reactions that occur on the surface of enzyme-coated particles \cite{Dey2015,Sch2015,Ma2016,Ma2016b,Zhao2018,toebes2019spatial,arque2019intrinsic,tang2020enzyme}. When an active particle participates in several chemical reactions, its self-propulsion may be fueled by multiple chemical species. Self-propelled active particles driven by multiple fuel species have been realized in the laboratory \cite{gao2013multi,Sch2017}. In contrast to active particles powered by the decomposition of hydrogen peroxide, enzyme-powered particles or motors are more desirable for biomedical applications due to their nontoxicity, versatility, and biofuel availability \cite{Sun2019,Chen2019,Yuan2021,Mathesh2020}.

In Section \ref{sec:model} a general formulation of the chemical reaction network and the Langevin equations of motion are presented in which both bulk reactions and surface reactions on the colloids are considered. We model the surface reaction as delta function sources or sinks but with general reaction kinetics. Bulk reactions are also included, which act to maintain the chemical system at a non-equilibrium state.  We then present three case studies of active colloids undergoing different reactions. In Section \ref{sec:single-reaction}, we show the clustering dynamics of a single type of colloid undergoing a single surface reaction. Because the bulk reaction is present, the chemical concentrations are screened and we observe localized clusters with phoretic attraction. We consider binary mixtures with different reaction kinetics in Sections \ref{sec:sequential} and \ref{sec:selkov}. In Section \ref{sec:sequential}, a sequential reaction is considered in which the product of the first reaction is the reactant of the second reaction in the sequence, and the first reaction occurs on the first type of colloid while the second reaction is on the second type of colloid. We show that this coupling induced by the sequential reaction gives rise to diverse clustering behavior. In Section \ref{sec:selkov} a binary mixture with a nonlinear autocatalytic reaction is considered before concluding the paper in  Section \ref{sec:conclusion}. 








\section{\label{sec:model} General formulation}
Consider a suspension of chemically active colloidal particles dispersed in an incompressible Newtonian fluid (solvent) of dynamic viscosity $\eta$. In addition to the $N$ active colloids, the suspension contains $n_s$ reacting solute species, $\bS = \{ S_j \mid j=1,\dots,n_s\}$. The local concentrations of $\bS$ at position $\bx$ and time $t$ are denoted as $\bc(\bx, t) = \{ c_j(\bx,t) \mid j=1,\dots,n_s\}$. The $n_s$ solute species form a chemical reaction network in which the solute $S_j$ can act as either  reactants ($R_j$) or  products ($P_j$) in potentially multiple chemical reactions. Possibly reversible chemical reactions, which were chosen from a set of $n_r$ such reaction schemes, occur on the surface of the $N$ active colloids. For a particular surface reaction indexed by $K$ ($K=1,\dots,n_r$), we may write the reaction scheme compactly as $\bR^K \rightleftharpoons \bP^K$. Here, the set of reactants is a subset of $\bS$,   $\bR^K=\{S_{j} \lvert j \in I^K\} \subset \bS $, where $I^K$ is the index set that contains the chemical species acting as reactants for the reaction indexed by $K$. Similarly, the set of products of reaction $K$ is $\bP^K=\{S_{j} \lvert j \in O^K\} \subset \bS$, where $O^K$ is the index set that contains the chemical species acting as products for the reaction indexed by $K$. The
local concentrations corresponding to these sets of species are similarly defined as $\bc_R^K(\bx, t)$ and $\bc_P^K(\bx, t)$. 

The local chemical concentration fields $\bc(\bx,t)$ evolve by the reaction-diffusion equations that account for reactions on the surfaces of the active particles as well as reactions in the fluid phase that both participate in the reaction network and serve to maintain the system in a nonequilibrium state so that active motion is possible. 
%
The reaction-diffusion equation is given by 
\begin{align}
\label{eq:general-reaction-diffusion}
\partial_t c_j(\bm{x},t) = &D_j \nabla^2 c_j - \sum_{K=1}^{n_r} \mathcal{R}^b_{j,K}(\bm{c}^K) \nonumber \\ 
  & +\sum_{K=1}^{n_r} \sum_{i_K=1}^{N_K}  \mathcal{R}^s_{j,K}(\bm{c}^K)\delta(\bm{x}-\bm{X}_{i_K}),
\end{align}
where $\bm{c}^K$ includes the concentrations of both the reactants $\bc_R^K(\bx, t)$ and  the products $\bc_P^K(\bx, t)$. 
The second term on the right side accounts for reactions in the bulk fluid phase that may help to maintain the nonequilibrium state of the system. The nonequilibrium state could also be established through chemostats at the boundaries. The third term accounts for surface reactions on the active particles and in writing this term we adopted a monopole approximation for the surface reactions, following previous studies \cite{Pohl_Stark_2014,pohl2015self,Liebchen2015, kondrat2016discrete,Liebchen2017}. The positions of the particle centers are denoted by $\bm{X}_{i_K}$. Note that the reaction indexed by $K$ occurs on the particle labeled $i_K$, and there are in total $N_K$ such particles.   As a result, in the third term on the right side of \eqref{eq:general-reaction-diffusion}, we first sum over all particles ($N_K$) that carry the reaction indexed by $K$ and then sum over all $n_r$ reactions.  In general, both the bulk and surface rates, $\mathcal{R}^b_{j,K}$ and $\mathcal{R}^s_{j,K}$, are nonlinear functions of the chemical concentrations, and both forward and reverse rates are included in the case of reversible reactions.  In \eqref{eq:general-reaction-diffusion}, if $c_j$ does not participate in the reaction indexed by $K$, the rate coefficients $\mathcal{R}^b_{j,K}$ and $\mathcal{R}^b_{j,K}$ are understood to be zero. For the remainder of this paper, we assume that all chemical species have the same molecular diffusivity $D$.

As a consequence of microscopic reversibility, the surface reaction rates also depend on the forces the colloids experience~\cite{GK17,GK19,DeCP22}, including external forces as well as colloid-colloid interactions arising from chemical gradients and direct interactions. The magnitude and sign of the effect depends strongly on the orientation of the Janus colloid relative to the force. In our simulations neither the forces a motor experiences nor its orientation vector are controlled; hence, in what follows, we may neglect this effect on the reaction rate (see ESI). We note that an external force applied to a collection of Janus colloids with controlled orientation vectors has been shown to change the net reaction rate of the collection.~\cite{GK-Research2020}

We consider monodisperse and spherical active particles of radius $a$. The linear equations of motion for the active particles may be written as
\begin{align}
\label{eq:linear-EOM}
\frac{\od}{\od t}\bX_{i_K}= V_{\rm sd} \big( \bm{c}^K \big) \bm{u}_{i_K} + \sum_{j =1}^{n_s} \mu_{i_Kj} \bm{\nabla} c_j +\frac{\bm{F}_{i_K}^P}{\zeta_{t}}+\bm{V}^B_{i_K}.
\end{align}
The first term on the right is the self-diffusiophoretic velocity directed along $\bm{u}_{i_K}$ that depends on the concentrations in the $K$ reaction, while the second term is the diffusiophoretic velocity due to the concentration gradients in the surroundings. The coefficient $\mu_{i_K j}$ represents the response of particle $i_K$ to the chemical species $j$; $\mu_{i_K j}$ may be zero for some species and take either sign ($+$ or $-$) depending on the physical context. In the third term $\bm{F}_{i_K}^P$ is the impulsive force on particle $i_K$ due to hard interactions with the other active particles, and $\zeta_{t}$ is the translational friction coefficient of the particles. The last term is the Gaussian, fluctuating Brownian velocity, which satisfies 
\begin{align}
    \langle \bm{V}^B_{i_K}(t) \rangle = \bm{0}\quad {\rm and}\quad \langle\bm{V}^B_{i_K}(t) \bm{V}^B_{i_K}(t^\prime ) \rangle =2 D_t \delta(t-t^\prime ) \bI, 
\end{align}
where $D_t$ is the translational diffusivity, and $\bI$ is the identity tensor. In writing \eqref{eq:linear-EOM}, hydrodynamic interactions among particles are neglected and the translational friction coefficient is taken to be $\zeta_{t} = 6 \pi \eta a$. The translational diffusivity is related to the thermal energy $k_BT$ via the relation $k_BT = \zeta_{t} D_t$.   We note that previous studies suggest that chemotactic forces may dominate the clustering dynamics while  hydrodynamic interactions play a less important role \cite{huang2017chemotactic}. 

The angular equations of motion for the active particles are given by 
\begin{align}
\label{eq:angular-EOM}
    \frac{\od }{\od t} \bu_{i_K}= \sum_{j=1}^{n_s} \gamma_{i_K j}\left(  \bu_{i_K}\times \bnabla c_j \right)\times \bu_{i_K} +  \bOmega^B_{i_K} \times \bu_{i_K}, 
\end{align}
where the signed constant $\gamma_{i_Kj}$ is the angular diffusiophoretic coefficient for species $j$, and $\bOmega^B_{i_K}$ is the fluctuating Brownian angular velocity given by 
\begin{align}
        \langle \bOmega^B_{i_K}(t) \rangle = \bm{0}\quad {\rm and}\quad \langle\bOmega^B_{i_K}(t) \bOmega^B_{i_K}(t^\prime ) \rangle =2 D_r \delta(t-t^\prime ) \bI,
\end{align}
where $D_r$ is the rotational diffusivity that satisfies $\zeta_rD_r = k_BT$, and $\zeta_{r} = 8\pi \eta a^3$ is the rotational friction coefficient. In \eqref{eq:angular-EOM}, notice that $\left( \bu_{i_K} \times \bnabla c_j \right)\times \bu_{i_K} = \left( \bI - \bu_{i_K}  \bu_{i_K}  \right)\cdot \bnabla c_j$.

Equations \eqref{eq:linear-EOM} and \eqref{eq:angular-EOM} are a generalization of the equations considered by previous studies that include one chemical species \cite{Pohl_Stark_2014,pohl2015self}. Modeling the chemical as a diffusing species in the presence of delta function sinks located at the particle centers, Pohl and Stark \cite{Pohl_Stark_2014,pohl2015self} showed that the colloids can establish  dynamic clustering states or collapse into a single cluster, depending on the phoretic parameters. In this model, the chemical concentration at a distance of $r$ away from a sink scales as $1/r$ (in 3D). As a result, if the phoretic interaction is purely attractive, this long-ranged interaction (the chemical gradient scales as $1/r^2$) ultimately leads to the aggregation of colloids into a single large cluster. In the current work, our aim is to demonstrate that  complex kinetics involving multiple chemical species can lead to more diverse clustering dynamics, which may be useful for the design of self-assembling active materials.

\section{Active colloids propelled by a single surface reaction \label{sec:single-reaction}}
We now consider the simplest case in which all particles participate in the same surface reaction, given by $S_1 \to S_2$. In the bulk, $S_1$ is replenished using the effective reaction $S_2 \to S_1$ so as to maintain a non-equilibrium state. We remark that the bulk reaction should not be treated as the reverse of the surface reaction but as a separate reaction with other participating chemical species. These additional chemical species are held at fixed concentrations by chemical reservoirs \cite{HSGK18}. As a result, the concentration of the reservoir species do not appear in the effective bulk reaction. This simple reaction scheme serves as an introduction to the general mechanisms of phoretic interactions and to the effects of chemical screening due to bulk reactions.

For this reaction scheme, the reaction-diffusion equation \eqref{eq:general-reaction-diffusion} becomes 
\begin{subequations}
\label{eq:c-eq-single}
\begin{align}
\label{eq:c1-eq-single}
    \frac{\partial c_1}{ \partial t} &=D \nabla^2 c_1 - \kappa \sum_{i=1}^N c_1 \delta(\bx - \bX_i) + k_b  c_2,  \\ 
    \label{eq:c2-eq-single}
    \frac{\partial c_2}{ \partial t} &=D \nabla^2 c_2 + \kappa \sum_{i=1}^N c_1 \delta(\bx - \bX_i) - k_b  c_2,
\end{align}
\end{subequations}
where $\kappa$ is an effective reaction rate for the surface reaction and $k_b$ is the bulk reaction rate. We note that in 3D, $\kappa$ is the surface rate coefficient (which has units of length/time) multiplied by the surface area; in 2D, it is multiplied by the circumference. Because both species have the same diffusivity, we obtain $\partial(c_1+c_2)/\partial t \equiv 0$. Suppose that initially $c_1+c_2=c_0=const.$, then the total concentration $c_0$ remains constant in time and homogeneous in space. Upon substituting $c_2 = c_0-c_1$ into \eqref{eq:c-eq-single}, one only needs to solve \eqref{eq:c1-eq-single}.  The equations of motion \eqref{eq:linear-EOM} and \eqref{eq:angular-EOM} are now written as 
\begin{subequations}
\label{eq:langevin-single}
    \begin{align}
    \label{eq:langevin-single-linear}
        \frac{\od }{\od t}\bX_i  &=\alpha \, c_1\, \bu_i  +\mu \,  \bnabla c_1  +\frac{\bm{F}_i^P}{\zeta_{t}}+ \bm{V}_i^B, \\ 
        \frac{\od }{\od t}\bu_i  &= \gamma\left(  \bu_{i}\times \bnabla c_1\right)\times \bu_{i} +    \bOmega^B_{i} \times \bu_{i},
    \end{align}
\end{subequations}
where the self-diffusiophoretic speed is taken to be linearly proportional to the local concentration, $V_{\rm sd} = \alpha c_1$. This form of a linear self-diffusiophoretic speed was considered in Ref.  \cite{Saha2014}. We have also assumed that all particles respond to the chemical gradient $\bnabla c_1$ in the same fashion and the subscripts of $\mu$  and $\gamma$ in \eqref{eq:linear-EOM} and \eqref{eq:angular-EOM} are dropped. 

In the presence of bulk ``refueling'' (i.e., $S_2 \to S_1$), the system \eqref{eq:c-eq-single} is chemically screened; the inverse screening length is 
\begin{align}
    \lambda = \sqrt{\frac{k_b}{D}}.
\end{align}
In the remainder of the paper, we focus on 2D systems with monodisperse disks of radius $a$. Equations \eqref{eq:c-eq-single} and \eqref{eq:langevin-single} are governed by the following dimensionless groups:
\begin{subequations}
\label{eq:non-dim}
    \begin{align}
    &\Lambda = \lambda a, \quad Da = \frac{\kappa }{4 \pi a D},\quad \hat{\alpha}= \frac{\alpha c_0 a}{D}, \\
    &\quad \hat{\mu} = \frac{\mu c_0}{D}, \quad \hat{\gamma} = \frac{\gamma c_0 a}{D}, \quad \frac{D_t}{D}.
\end{align}
\end{subequations}

Here, $Da$ is a Damk\"{o}hler number that compares the surface reaction rate to the rate of diffusion. We note that $a^2 D_r/D_t = 3/4$ for isolated spheres of radius $a$. Since the chemical species often diffuse faster than the colloids, in the remainder of the paper, we fix $D_t/D$ to be $0.1$ and focus on the variation of the other dimensionless parameters. Finally, the area fraction of the particles is important and denoted as $\phi$.

\begin{figure}[h]
\centering
  \includegraphics[width=0.45\textwidth]{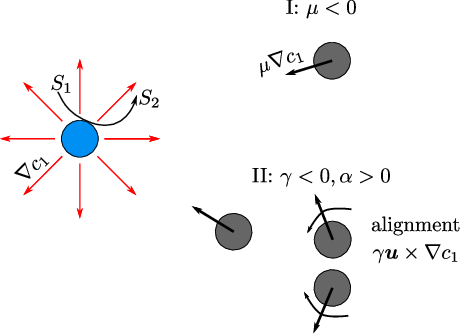}
  \caption{Schematic of effective attractions between two active particles mediated by chemical interactions. The red arrows surrounding the blue particle indicate the concentration gradient $\bnabla c_1$. (I): For $\mu <0$, the gray particle moves towards the blue particle due to the diffusiophoretic velocity in the presence of the gradient, $\mu\bnabla c_1$. (II): For $\gamma <0$ and $\alpha >0$, the gray particles are first rotated ($\gamma \bm{u}\times \bnabla c_1$) and then can swim towards the blue particle with their self-diffusiophoretic velocity, $\alpha c_1 \bm{u}$. }
  \label{fig:attraction-schematic}
\end{figure}

The single surface reaction model specified by \eqref{eq:c-eq-single} and \eqref{eq:langevin-single} can be treated as a generalization of previous models \cite{Pohl_Stark_2014,pohl2015self,Liebchen2017}. More specifically,  the swim speed (the first term on the right side of \eqref{eq:langevin-single-linear}) in these previous models is taken to be a constant and does not explicitly depend on the local chemical concentration $c_1$. Furthermore, in Pohl and Stark \cite{Pohl_Stark_2014,pohl2015self}, the bulk refueling is absent; in \citet{Liebchen2017}, $\mu \equiv 0$, and the active particles do not exhibit translational diffusiophoretic motion as a result of the gradient $\bnabla c_1$ induced by other particles. 

\begin{figure*}[t]
 \centering
 \includegraphics[width=\textwidth]{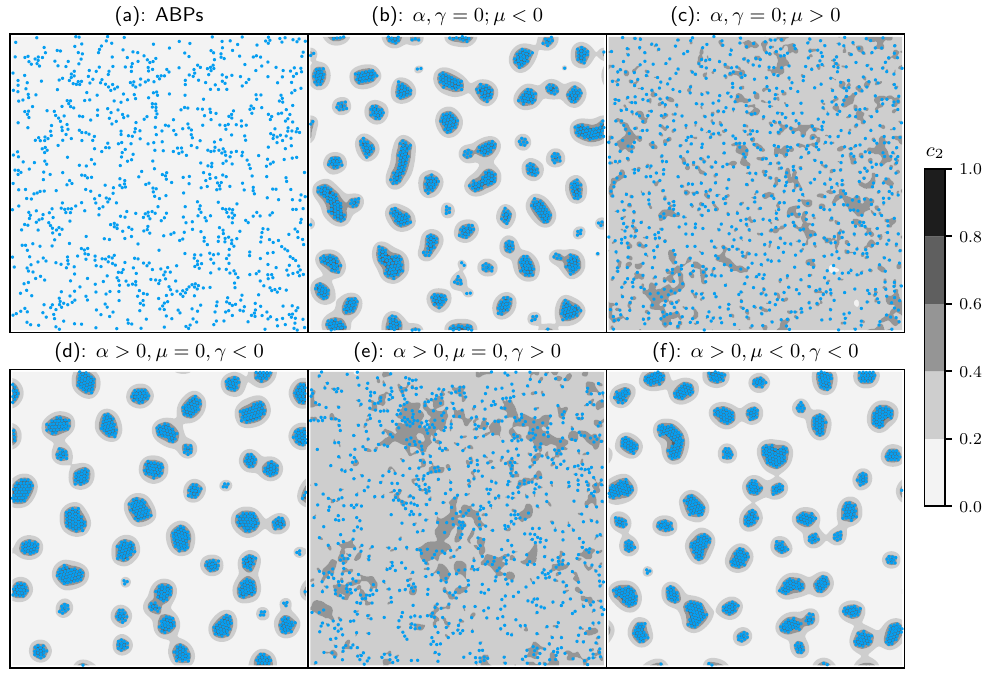}
 \caption{Representative snapshots of active particles at area fraction $\phi=0.1$. The red background is the concentration field of the product species $c_2$. (a): Snapshot of active Brownian particles, $Da, \hat{\mu}, \hat{\gamma}=0$ and $\hat{\alpha} = 10$. In this case, the concentration remains a constant in space and time, $c_1 \equiv c_0$ and $c_2 \equiv 0$, because initially $c_1 = c_0$. (b): Clustering of active particles due to phoretic attraction (mechanism I in Fig. \ref{fig:attraction-schematic}), $\hat{\alpha}, \hat{\gamma}=0; \hat{\mu} = -10$. (c): Homogeneous state  in the presence of phoretic repulsion,$\hat{\alpha}, \hat{\gamma}=0; \hat{\mu} = 10$. (d): Clustering of active particles due to phoretic attraction (mechanism II in Fig. \ref{fig:attraction-schematic}), $\hat{\alpha}=10, \hat{\mu}=0, \hat{\gamma}=-10$. (e): Homogeneous state in the presence of phoretic repulsion, $\hat{\alpha}=10, \hat{\mu}=0, \hat{\gamma}=10$.  (f): Clustering of active particles in the presence of both phoretic attraction mechanisms, $\hat{\alpha}=10, \hat{\mu}=-10, \hat{\gamma}=-10$. For all panels, $\Lambda = 0.25$. For (b)-(f), $Da=1/(2\pi) \approx 0.16$. }
 \label{fig:snapshots-single-rxn}
\end{figure*}

In our colloid model, which is similar to that of Saha, et al.~\cite{Saha2014}, there are two phoretic mechanisms that lead to effective attraction between particles (see Fig. \ref{fig:attraction-schematic}). The first phoretic attraction mechanism originates from the phoretic translation of a particle (gray) in the presence of the chemical gradient induced by another active particle (blue). As shown in Fig. \ref{fig:attraction-schematic}, the surface reaction on the blue particle induces a radial concentration gradient in the vicinity. For $\mu <0$, the concentration gradient will induce a translational velocity, $\mu \bnabla c_1$,  that directs the second particle nearby towards the blue particle. We note that for $\mu >0$, the effective interaction becomes repulsive. The second mechanism for attraction results from the combined effect of self-diffusiophoretic swimming, $\alpha c_1 \bu$,  and phoretic alignment, $\gamma \bu \times \bnabla c_1$. In  the radial concentration gradient, the gray particles experience a diffusiophoretic torque that tends to align the swimming direction $\bu$ against the gradient $\bnabla c_1$ provided that $\gamma <0$. Once aligned, the gray particle can swim towards the blue particle via self-diffusiophoretic translational motion if $\alpha >0$. Again, if $\gamma >0$, the effective interaction becomes repulsive. Because of the conservation, $c_0 = c_1+c_2$, we have $\bnabla c_1 = - \bnabla c_2$; one can then reformulate \eqref{eq:langevin-single} in terms of $c_2$, in which case the sign of $\mu$ and $\gamma$ needs to be reversed in order to have attractive interactions.

Typical snapshots of the system at long times with an area fraction $\phi =0.1$ are shown in Fig. \ref{fig:snapshots-single-rxn}. In the absence of surface reaction ($Da =0$), the chemical concentration remains a constant in space and time if initially $c_1 = c_0 = const.$ As a result, the system reduces to the bare dynamics of active Brownian particles with hard-sphere interactions; the configuration of the system is shown in Fig. \ref{fig:snapshots-single-rxn}(a). At this low density, motility-induced phase separation \cite{Fily_Marchetti_12,Redner_2013,Cates2015,Takatori_2015,Solon_2018,Hermann_2019,Omar_2023} is absent and the system remains in a homogeneous state with visible density fluctuations. In Fig. \ref{fig:snapshots-single-rxn}(b), the self-diffusiophoretic swimming is turned off ($\alpha=0$), and particles form localized clusters due to phoretic attraction induced by the chemical gradient (mechanism I in Fig. \ref{fig:attraction-schematic}). If one reverses the sign of $\mu$ ($-$ to $+$), the effective interaction becomes repulsive, and particles remain in a homogeneous state (see Fig. \ref{fig:snapshots-single-rxn}(c)).

In Fig. \ref{fig:snapshots-single-rxn}(d), $\mu=0$ and $\gamma <0$, particles form clusters due to the second mechanism of phoretic attraction (see Fig. \ref{fig:attraction-schematic}). In contrast to  Fig. \ref{fig:snapshots-single-rxn}(a), we see that the rotation and alignment from phoretic interactions allow one to turn the random distribution of active Brownian particles into localized clusters.  If one reverses the sign of $\gamma$ from negative to positive, particles experience effective repulsion and do not form clusters (see Fig. \ref{fig:snapshots-single-rxn}(e)). Lastly, in Fig. \ref{fig:snapshots-single-rxn}(f), both mechanisms of attraction are present and one observes cluster formation.  Because the chemical system is screened, the clusters are localized and one does not observe the aggregation of all particles into a single cluster (cf. \citet{Pohl_Stark_2014}). We note that with screening large-scale clusters can be obtained for higher area fractions.

\section{Binary mixture with sequential reactions}
\label{sec:sequential}

\begin{figure}
\centering
  \includegraphics[width=0.28\textwidth]{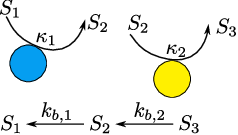}
  \caption{Schematic of the sequential reaction scheme. The surface reaction $S_1 \to S_2$ occurs on the first type of particles while $S_2 \to S_3$ occurs on the second type. }
  \label{fig:sequential-schematic}
\end{figure}

We now turn our attention to binary mixtures. The system of active particles is classified into two types. The surface reaction $S_1 \to S_2$ occurs on the surface of type 1 particles. For type 2 particles, we have the surface reaction $S_2 \to S_3$. Similar to the previous section, bulk reactions $S_3 \to S_2$ and $S_2\to S_1$ (see Fig. \ref{fig:sequential-schematic} for a schematic) are incorporated to maintain the system at a nonequilibrium state (see Fig. \ref{fig:sequential-schematic}). Similar to the reactions considered in the previous section, the bulk reactions are effective reactions that are maintained out of equilibrium using chemical reservoirs of other species. The reaction-diffusion system governing the concentration $c_i$, $i=1,2,3$, is given by 
\begin{subequations}
    \begin{align}
        \label{eq:c1-PDE-sequence}
    \frac{\partial c_1}{\partial t} =& D \nabla^2 c_1 - \kappa_1 \sum_{i\in \sigma_1}c_1\delta\left(\bx - \bX_i\right)  + k_{b,1} c_2, \\ 
    \label{eq:c2-PDE-sequence}
    \frac{\partial c_2}{\partial t} = &D \nabla^2 c_2 + \kappa_1 \sum_{i\in \sigma_1} c_1\delta\left(\bx - \bX_i\right)  - k_{b,1} c_2 \nonumber \\
    & - \kappa_2 \sum_{j \in \sigma_2}c_2\delta\left(\bx - \bX_j\right)  + k_{b,2} c_3 ,\\
    \label{eq:c3-PDE-sequence}
    \frac{\partial c_3}{\partial t} =& D \nabla^2 c_3 + \kappa_2 \sum_{j \in \sigma_2}c_2\delta\left(\bx - \bX_j\right)  - k_{b,2} c_3 .
    \end{align}
\end{subequations}
In the above, $\sigma_1 = \{i \,\lvert \,i \in \mathbb{Z}, 1 \leq i \leq N_1 \}$ denotes the index set of $N_1$ type 1 particles. There are $N-N_1$ type 2 particles denoted by the set $\sigma_2 = \{i \,\lvert \, i \in \mathbb{Z}, N_1 < i \leq N \}$.  The rate coefficients,  $\kappa_1$ and  $\kappa_2$,  are the surface rate of the first and second reaction, respectively; $k_{b,1}$ and $k_{b,2}$ are the bulk reaction rate coefficients of the reactions $S_2 \to S_1$ and $S_3 \to S_2$, respectively.  Adding \eqref{eq:c1-PDE-sequence}--\eqref{eq:c3-PDE-sequence}, we see that the chemical species satisfy the conservation condition, $c_1+c_2+c_3 = c_0=const.$  The active velocity for the type $1$ particle with index $i$ is given by $\alpha_1 c_1 \bu_i + \mu_{11} \bnabla c_1 + \mu_{12} \bnabla c_2$; for the type 2 particle with index $j$, we have the active velocity given by $\alpha_2 c_2 \bu_j + \mu_{21} \bnabla c_1 + \mu_{22} \bnabla c_2$. Because $\bnabla c_3 = -\left(  \bnabla c_1 + \bnabla c_2\right)$, the term $\mu_{13} \bnabla c_3$ is not needed and has been absorbed into the coefficients $\mu_{11}$ and $\mu_{12}$.  The active velocity for type 2 particles is treated similarly. The active angular velocities for type 1 and 2 are, respectively, $ \bu_{i}\times \left( \gamma_{11} \bnabla c_1  + \gamma_{12} \bnabla c_2\right)$ and $ \bu_{j}\times \left( \gamma_{21} \bnabla c_1  + \gamma_{22} \bnabla c_2\right)$.  The fluctuating and hard-sphere terms are the same as those given in \eqref{eq:langevin-single}.

\begin{figure*}[t]
 \centering
 \includegraphics[width=\textwidth]{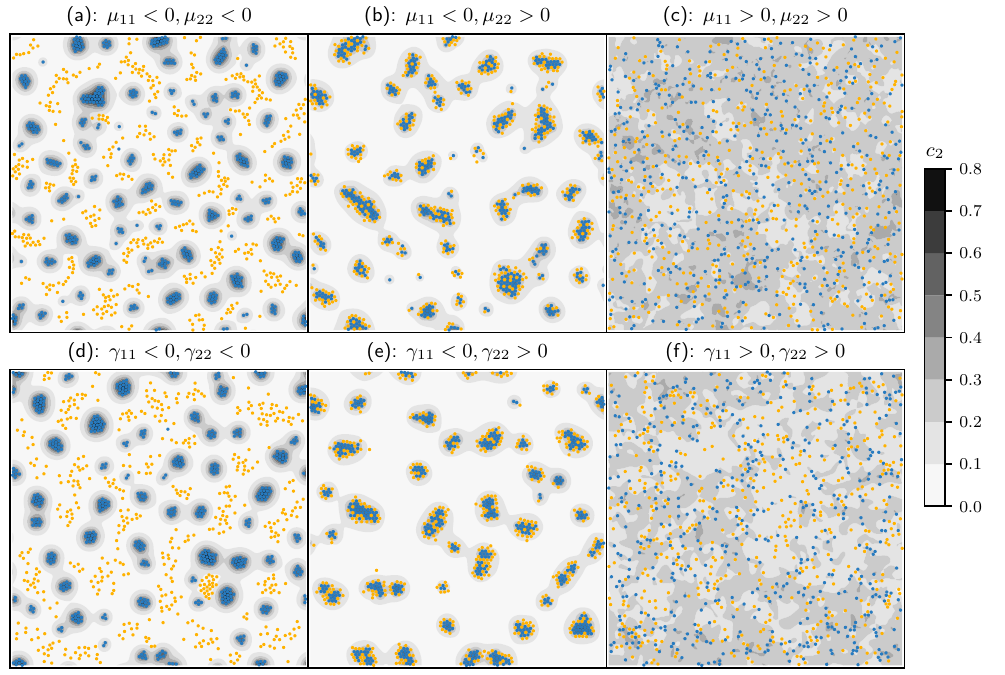}
 \caption{Representative snapshots of binary mixtures of active particles undergoing sequential surface reactions at total area fraction $\phi=0.1$. Type 1 particles are shown in blue and type 2 are orange. The background is the concentration field of the intermediate species $c_2$. For all simulations, the particles are initialized with random positions and orientations, the initial concentrations are $c_1 = c_0$, $c_2=c_3=0$,  $\Lambda_1 =\Lambda_2=0.25$, $Da_1=Da_2=1/(2\pi) \approx 0.16$, $\hat{\mu}_{12} = \hat{\mu}_{21}=0$, $\hat{\gamma}_{12} = \hat{\gamma}_{21}=0$, and the number of type 1 particles are the same as 2. For (a)--(c),$\hat{\alpha}_{1}=\hat{\alpha}_{2}=\hat{\gamma}_{11}=\hat{\gamma}_{22}=0$; for (d)--(f), $\hat{\mu}_{11}=\hat{\mu}_{22}=0$, and $\hat{\alpha}_1=\hat{\alpha}_2=10$.
(a): Type 1 particles form clusters while type 2 are dispersed, $\hat{\mu}_{11}=\hat{\mu}_{22}=-20$.
(b): Heterogeneous clusters are observed that contain both types of particles, $\hat{\mu}_{11}=-20, \hat{\mu}_{22}=20$.
(c): Both 1 and 2 are dispersed, $\hat{\mu}_{11}=\hat{\mu}_{22}=20$.
(d): Type 1 particles form clusters while type 2 are dispersed, $\hat{\gamma}_{11}=\hat{\gamma}_{22}=-20$.
(e): Heterogeneous clusters are observed that contain both types of particles, $\hat{\gamma}_{11}=-20, \hat{\gamma}_{22}=20$.
(f): Both 1 and 2 are dispersed, $\hat{\gamma}_{11}=\hat{\gamma}_{22}=20$.
}
 \label{fig:snapshots-sequential}
\end{figure*}

In this binary model, the chemical coupling between the dynamics of the two types of particles is through the intermediate species $S_2$. The equations are non-dimensionalized following the scheme given in \eqref{eq:non-dim}.  In contrast to the single reaction case considered in the previous section, we now have an even larger set of dimensionless parameters. For simplicity, we consider the case in which the cross terms $\mu_{12}, \mu_{21}, \gamma_{12}$ and $\gamma_{21}$ are zero. We focus on the effects of $\mu_{11}, \mu_{22}, \gamma_{11}$ and $\gamma_{22}$ on the clustering behavior.

\begin{figure}[ht]
 \centering
 \includegraphics[width=3in]{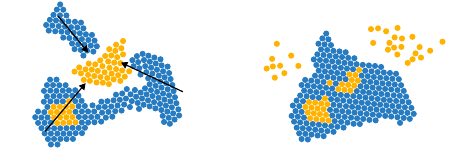}
 \caption{Illustration of the 1-1 attraction induced trapping of type 2 particles. Left: the blue clusters are attracted towards each other, their motion are indicated by the black arrows. Right: as the blue clusters aggregate, some gold particles escape from the two open channels shown on the left while the rest are enclosed by the large cluster. The schematic is based on the simulation shown in Fig. 9 of the ESI. 
 \label{fig:trapping-process}
 }
\end{figure}

In Fig. \ref{fig:snapshots-sequential}, we show typical simulation snapshots of the system at long times with a total (all particles) area fraction $\phi=0.1$. The blue particles in Fig. \ref{fig:snapshots-sequential} denote type 1 while the orange are type 2.  When $\mu_{11}<0$, type 1 particles tend to aggregate because they consume $S_1$ (sinks of $S_1$), which induces a phoretic attraction. This aggregation of type 1 particles is shown in Fig. \ref{fig:snapshots-sequential}(a) and (b). In Fig. \ref{fig:snapshots-sequential}(a), $\mu_{22} <0$, type 2 particles are repelled from sources of $S_2$, i.e., type 1 particles. Since $\mu_{22} <0$, type 2 particles are also attracted to sinks of $S_2$, or type 2 particles. Because $S_2$ is only produced by type 1 particles (and in the fluid from $S_3$) and the reaction is screened, the concentration of $S_2$ ($c_2$) is very small at locations of type 2 particles (see the background contour in Fig. \ref{fig:snapshots-sequential}(a)). In addition, the bulk reaction $S_2 \to S_1$ diminishes the concentration of $S_2$ near type 2 particles. This, in turn, reduces the production of $S_3$ (on type 2 particles) and hence $S_2$ by the bulk reaction $S_3 \to S_2$.  As a result, the attraction among type 2 particles is weak and one observes that they remain loosely dispersed away from islands of type 1 particles. Effectively, in Fig. \ref{fig:snapshots-sequential}(a), the binary system exhibits a phoretic attraction among type 1 particles and type 2 particles are repelled from type 1. This phoretic repulsion constitutes a non-reciprocal chemical interaction because type 2 are repelled from type 1, but not vice versa.  For weaker screening, or larger screening length, the size of type 1 particle clusters is larger. In this case, we observe that type 2 particles are pushed away from type 1 clusters and appear to be more closely aggregated (see ESI, Fig. 1). As multiple type 1 clusters move towards each other due to longer range attraction, they may squeeze a collection of type 2 particles together and enclose them as part of the merged cluster (see Fig. \ref{fig:trapping-process}).

When the sign of $\mu_{22}$ is switched from negative to positive (Fig. \ref{fig:snapshots-sequential}(a) $\to$ Fig. \ref{fig:snapshots-sequential}(b)), type 1 particles become attractive to type 2 and we observe heterogeneous clusters consisting of both types of particles. The clustering dynamics in Fig. \ref{fig:snapshots-sequential}(b) is subtle and warrants a more detailed consideration. In particular, we notice that  type 2 particles are often on the periphery of the clusters while type 1 particles occupy the center.  Because type 1 particles consumes $S_1$ and $\mu_{11}<0$, they tend to move towards particles of the same type and form clusters of type 1. Once a type 1 cluster of several particles is formed, a noticeable density gradient of $S_2$ is established locally. Since $\mu_{22} >0$, type 2 particles tend to move towards sources of $S_2$, which are the localized clusters of type 1 particles. As a result, type 2 particles are attracted to type 1 from the term $\mu_{22} \bnabla c_2$ but type 1 particles are not attracted to type 2. This non-reciprocal attraction ultimately leads to clusters in which type 2 particles are attached to the periphery. We also note that type 2 particles do not form clusters on their own because they consume $S_2$ but are attracted to sources of $S_2$.

In Fig. \ref{fig:snapshots-sequential}(c), $\mu_{11}>0$, which implies that type 1 particles are repelled from sinks of $S_1$. In other words, they exhibit effective repulsion and do not form clusters or establish a noticeable concentration gradient of $S_2$. In the absence of $\bnabla c_2$, type 2 particles will also become dispersed since there are no type 1 clusters to which they are attracted. As a result, we see that both types of particles are dispersed. 

For Figs. \ref{fig:snapshots-sequential}(a)--(c), the phoretic interactions result from the type I mechanism outlined in Fig. \eqref{fig:attraction-schematic}, though with more than one chemical species. Just like the single reaction case considered in the previous section, one can achieve mixture clustering dynamics similar to those shown in Figs. \ref{fig:snapshots-sequential}(a)--(c) using the type II mechanism. These results are shown in  Figs. \ref{fig:snapshots-sequential}(d)--(f), where the type I mechanism is turned off. That is, $\mu_{ij} =0$, $i,j = 1,2$, and $\alpha_1, \alpha_2, \gamma_{11}$ and $\gamma_{22}$ are nonzero. In this case, particles self-propel with their self-diffusiophoretic velocities and are rotated to align with or against density gradients depending on the sign of $\gamma_{ij}$. In Fig. \ref{fig:snapshots-sequential}(d), $\gamma_{11}, \gamma_{22} <0$, and we observe clustering of type 1 particles  as a result of their phoretic attraction; similar to Fig. \ref{fig:snapshots-sequential}(a), type 2 particles are dispersed in the bulk.  In Fig. \ref{fig:snapshots-sequential}(e), $\gamma_{11}<0$ and  $\gamma_{22} >0$, type 2 particles are attracted to type 1 clusters. When $\gamma_{11}, \gamma_{22}>0$, both type 1 and type 2 are dispersed, as shown in \ref{fig:snapshots-sequential}(f).

In Fig. \ref{fig:snapshots-sequential}(b) and (e), we see that there are type 2 particles trapped in the middle of the clusters. This occurs because two small type 1 clusters with type 2 particles attached to their periphery can merge into a single cluster as a result of their phoretic attraction, which leaves some type 2 particles trapped in the middle of the new cluster. To show this merging process, in Fig. \ref{fig:snapshots-sequential-time} we plot the system snapshots at different times. In contrast to the simulation of Fig. \eqref{fig:snapshots-sequential}(b), the phoretic coefficient $\hat{\mu}_{22}$ in the simulation of Fig. \ref{fig:snapshots-sequential-time} has a delayed activation. That is, initially $\hat{\mu}_{22} =0$, once the system establishes type 1 clusters (Fig. \ref{fig:snapshots-sequential-time}(b)), $\hat{\mu}_{22}$ is activated ($\hat{\mu}_{22}=-20$). Once $\hat{\mu}_{22}$ is turned on, type 2 particles move towards the already established type 1 clusters. As the system evolves, clusters will tend to merge if they are close enough to ``sense'' the chemical gradients from each other. In  Fig. \ref{fig:snapshots-sequential-time}(c), the numbered circles 1--3 highlight the merging process of two clusters into one. In circle 4, the two small clusters have already merged and we observe a line of trapped type 2 particles. (Lastly, we note that an earlier limited study using particle-based simulations of a binary mixture of sphere-dimer motors has shown that sequential kinetics also leads to distinctive cluster formation in this system.~\cite{CRRK14})

\begin{figure}[ht]
 \centering
 \includegraphics[width=\columnwidth]{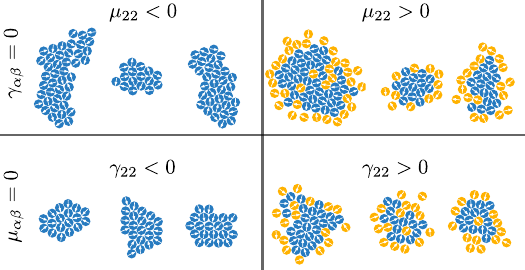} 
 \caption{Particle orientations in typical clusters. The particle orientation vector is denoted by the white arrow. The left and right panels of the top row correspond to Fig. \ref{fig:snapshots-sequential}(a) and (b), respectively. For the top row, $\gamma_{\alpha\beta}=0$.  The left and right panels of the bottom row correspond to Fig. \ref{fig:snapshots-sequential}(d) and (e), respectively. For the bottom row, $\mu_{\alpha\beta}=0$. 
 \label{fig:orientations}
 }
\end{figure}

While the heterogeneous clusters formed due to type I (Fig. \ref{fig:snapshots-sequential}(b)) and type II (Fig. \ref{fig:snapshots-sequential}(e)) mechanisms are similar, they can be distinguished by the orientational order within clusters. In Fig.~\ref{fig:snapshots-sequential}(b), particles aggregate due to attraction, i.e., they move towards each other with induced phoretic translational velocities. When they form a cluster, the particles are held together by attraction. The particle orientation vector is not affected by the attraction and remains uniformly randomly distributed in clusters (see the top row of Fig. \ref{fig:orientations}). For type II mechanism shown in Fig. \ref{fig:snapshots-sequential}(e), particles reorient due to phoretic alignment. In order to form a cluster, particles have to swim towards each other. In order to maintain a cluster, particles on the outer shells have to continue pushing towards the interior of the cluster (see the bottom row of Fig. \ref{fig:orientations}). A particle at the outer shell with an orientation vector pointing away from the cluster simply swims away.   

\begin{figure*}[t]
 \centering
 \includegraphics[width=6in]{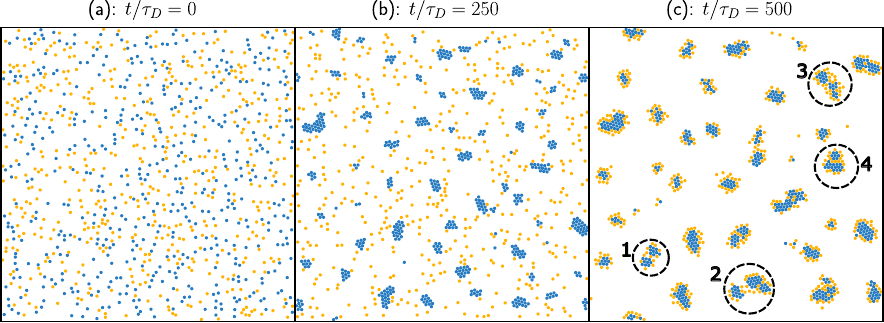}
 \caption{Snapshots of the binary mixture undergoing sequential surface reactions with total area fraction $\phi=0.1$ at different times: (a) $t/\tau_D=0$, (b) $t/\tau_D=250$, and (c) $t/\tau_D=500$,  where $\tau_D = a^2/D$ is the diffusive time scale. In terms of the bulk rate, we have for (b) $k_{b,1}t=15.625$,  and for (c) $k_{b,1}t=31.25$. Initially, $\hat{\mu}_{22} =0$; after (b), $\hat{\mu}_{22}$ is activated, $\hat{\mu}_{22} = -20$. All other parameters are the same as Fig. \ref{fig:snapshots-sequential}(b).  Numbered circles 1--3 show the merging process of two clusters into one while circle no. 4 shows a merged cluster. 
}
 \label{fig:snapshots-sequential-time}
\end{figure*}

\begin{figure}[th]
 \centering
 \includegraphics[width=3in]{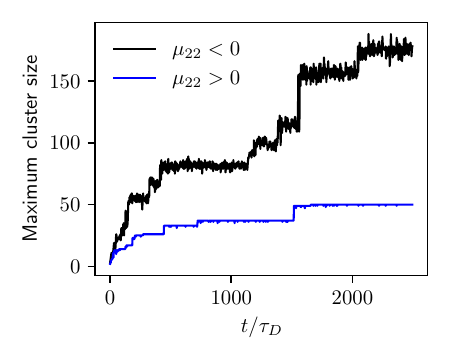}
 \caption{Maximum cluster size, defined as the number of particles in a cluster, in the system as a function of time. The black line, $\mu_{22}<0$, corresponds to the simulation presented in Fig. \ref{fig:snapshots-sequential}(a).  The blue line, $\mu_{22}>0$, corresponds to the simulation presented in Fig. \ref{fig:snapshots-sequential}(b). In counting the number of particles in a cluster, both types of particles are included.
 \label{fig:sequential-cluster-size}
 }
\end{figure}

In Fig.~\ref{fig:sequential-cluster-size}, we plot the maximum cluster size in the system as a function of time for simulations corresponding to Figs. \ref{fig:snapshots-sequential}(a) and \ref{fig:snapshots-sequential}(b). The maximum cluster size is defined as the total number of particles, regardless of their type,  in a single cluster. The system first establishes small clusters and then these clusters may merge due to attraction (see also Fig. \ref{fig:snapshots-sequential-time} and ESI). Such merging events are signified by the jumps in the cluster size curves shown in Fig.~\ref{fig:sequential-cluster-size}. The maximum cluster size for $\mu_{22}<0$ (black) is larger than that for $\mu_{22}>0$ (blue) because type 2 particles are attracted to type 1 particles or clusters when $\mu_{22}<0$. The fluctuations in the curves indicate that particles can randomly attach to or detach from a cluster as a result of Brownian motion. The jumps in the cluster size curve are more significant if two large clusters merge into a single one; this can occur at higher particle densities or for weaker chemical screening (see Figs. 1 and 6 in the ESI).

\begin{figure}[th]
 \centering
 \includegraphics[width=\columnwidth]{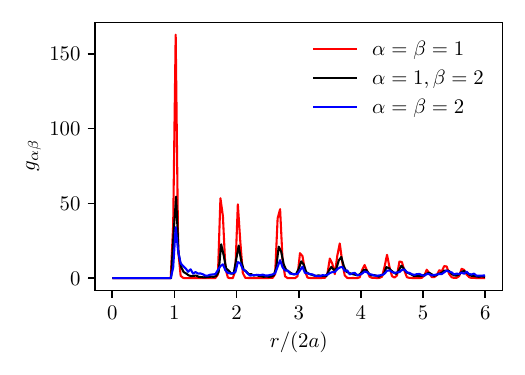}
 \caption{The self and cross radial distribution functions (RDFs) of the binary mixture that forms heterogeneous clusters. The RDFs are computed using a single frame of the system snapshot at $t/\tau_D=2497.5$. The system parameters are the same as those in Fig. \ref{fig:snapshots-sequential}(b). The subscripts $\alpha, \beta =1,2$ denote the particle type. The RDFs are computed with 150 bins for $r/(2a) \in [0, 6]$, where $r$ is the radial distance. 
 \label{fig:rdf}
 }
\end{figure}

For the system parameters used in Figs.~\ref{fig:snapshots-sequential}(b) and \ref{fig:snapshots-sequential-time}, the binary mixture forms heterogeneous clusters with a layering structure. In Fig.~\ref{fig:rdf} we plot the self and cross radial distribution functions (RDFs) for a snapshot taken at $t/\tau_D=2497.5$.  As can be seen in Fig.~\ref{fig:snapshots-sequential}(b), the clusters tend to approach a hexagonal packing with more loosely attached particles in the exterior. Because of the layering structure, the 1-1 RDF ($g_{11}$) has the largest peaks, which is followed by the 1-2 RDF ($g_{12}$). The 2-2 RDF has the lowest peaks because there are no attraction among type 2 particles; the peaks in $g_{22}$ largely results from the trapped particles (the layers) between type 1 particles.

We note that various system parameters may be used to tune the dynamics and self-organization of the binary mixture. In Fig.~\ref{fig:snapshots-sequential}(b) and \ref{fig:snapshots-sequential-time}, the phoretic coefficients $\mu_{11}$ and $\mu_{22}$ have the same magnitude but opposite sign, $\hat{\mu}_{22} = -\hat{\mu}_{11}$. By varying the relative magnitude $\mu_{22}/|\mu_{11}|$ while keeping the signs fixed, the system can be placed in different states (see ESI, Fig. 3). If $\mu_{22}$ is small, type 2 particles experience a weak attraction to type 1 clusters; as a result, the clusters are mostly of type 1 particles with many type 2 particles dispersed in bulk. For intermediate values of $\mu_{22}$, type 2 particles are attached to the clusters. If $\mu_{22}$ is large,  type 1 clusters can be destroyed by type 2 particles as a result of non-reciprocal attraction. Whenever a blue cluster forms, it attracts type 2 particles, which then move towards the cluster and can push through it; as a result, the clusters are destroyed.  Another parameter that may be accessible is the screening length, which can be modified by tuning the bulk reaction rate (see ESI, Fig. 1). If the screening length is much smaller than the particle size, no clusters can form since other particles do not sense the chemical gradient.  For large screening length (or weak screening), the system can form larger clusters even at low particle densities.

\begin{figure}[th]
 \centering
 \includegraphics[width=3in]{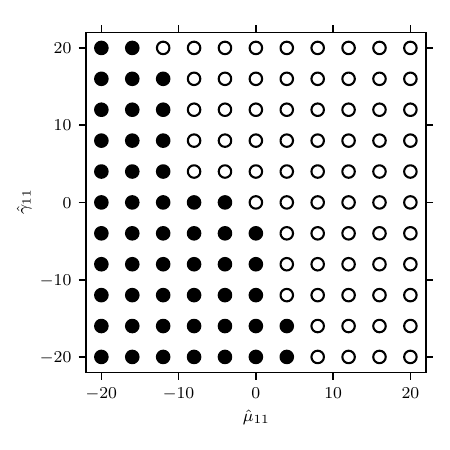}
 \caption{Clustering phase diagram for type 1 particles in the binary mixture undergoing the sequential reaction. Filled circles denote the clustering state while open circles imply that type 1 particles do not form clusters. The system parameters are $\Lambda_1 =\Lambda_2=0.25$, $Da_1=Da_2=1/(2\pi) \approx 0.16$, $\hat{\mu}_{12} = \hat{\mu}_{21}=\hat{\mu}_{22}=0$, $\hat{\gamma}_{12} = \hat{\gamma}_{21}= \hat{\gamma}_{22}=0$, $\hat{\alpha}_1=\hat{\alpha}_2=2$, and the total area fraction is $\phi=0.1$. 
 \label{fig:phase-plot}
 }
\end{figure}

Instead of varying the phoretic coefficients for different colloidal types, one can also consider different combinations of translational and rotational phoretic coefficients for the same particle type.  We present a phase diagram in Fig. \ref{fig:phase-plot} for the clustering of type 1 particles by varying $\mu_{11}$ and $\gamma_{11}$ while keeping all other parameters fixed.  In the lower left corner of the diagram, both $\mu_{11}$ and $\gamma_{11}$ are negative, which gives rise to attractive interactions from both type I and II mechanisms. As a result, in this region type 1 particles form clusters (filled circles) . When both $\mu_{11}$ and $\gamma_{11}$ are positive (upper right corner), no cluster is formed (open circles). Near the phase boundary, for example, $\hat{\gamma}_{11}=-20$ and $\hat{\mu}_{11}=4$,  the translational and rotational phoretic effects are in competition. For this particular case, the system can establish clusters; if $\hat{\mu}_{11}$ is increased, the clusters vanish and the system moves into a dispersed state.

In this section, we have illustrated the dynamics of a binary mixture with sequential reactions using three chemical species $S_1$, $S_2$ and $S_3$. The three chemical species form a closed chemical cycle (or loop) in the sense that all species are produced and consumed within the system, albeit at different ``sites''. As sketched in Fig.~\ref{fig:sequential-schematic}, $S_1$ is produced in the bulk and consumed on type 1 particles; $S_2$ is produced and consumed in the bulk and on particles; $S_3$ is produced on type 2 particles and consumed in the bulk. (The cycle can be ``short-circuited'' since $S_2$ produced in the bulk can react with type 2 particles.) In Refs. \cite{Golestanian2023PRL,Golestanian2023NatCommun}, colloidal mixtures that participate in reactions that form a closed cycle are considered where all chemical reactions occur on particle surfaces and chemical species freely diffuse in bulk. In contrast, the chemical cycle in our system is formed only when the particles  interact with the surrounding chemical environment (The chemical reactions on the particles do not form a cycle.). Notably,  the bulk reactions in the environment introduces chemical screening, which fundamentally changes the chemical interaction from long-ranged (pure diffusion) to short-ranged (screening). By tuning the chemical environment, i.e., the bulk reaction rate, one can control the system dynamics via the screening lengths. Because the system in general has multiple screening lengths, they can be designed to be different, thus may give rise to more complex dynamics. 

\section{Binary mixture with a nonlinear autocatalytic reaction}
\label{sec:selkov}
\begin{figure}
\centering
  \includegraphics[width=0.28\textwidth]{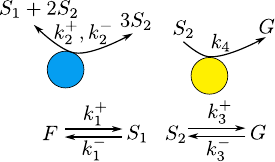}
  \caption{Schematic of the Selkov reaction scheme. The autocatalytic  reaction  occurs on the first type of particle while type 2 particles consume $S_2$ with a first-order reaction. Reversible reactions that produce or consume $S_1$ and $S_2$ are present in the bulk and the species $F$ and $G$ are maintained by reservoirs and their concentrations do not enter the reaction kinetics. }
  \label{fig:selkov-schematic}
\end{figure}

We now consider a binary mixture with nonlinear reaction kinetics based on the Selkov model \cite{selkov1968}. This model has its origin as simple description of a step in glycolsis involving the phosphofructokinase enzyme that converts fructose-6-phosphate and adenosine triphosphate (ATP) to fructose-1,6-bisphosphate and adenosine diphosphate (ADP). The reversible version of this model~\cite{richter1981control} comprises the following reactions: on enzyme-coated type 1 particles we have $S_1 +2S_2 \xrightleftharpoons[k_2^-]{k_2^+} 3S_2 $. In the bulk fluid, we have $F \xrightleftharpoons[k_1^-]{k_1^+} S_1$ and $S_2  \xrightleftharpoons[k_3^-]{k_3^+}  G$, where $F$ and $G$ are held fixed by reservoirs and their concentrations do not enter the reaction kinetics directly. On type 2 particles, we have $S_2 \xrightarrow{k_4} G$ (see Fig. \ref{fig:selkov-schematic} for a schematic). The destruction of $S_2$ on type 2 is not part of the Selkov model but is introduced to consume excess $S_2$ produced by the autocatalytic reaction on type 1. In the Selkov model $S_1$ and $S_2$ signify ATP and ADP, respectively, but here we regard it as a representative model for nonlinear autocatalytic kinetics.

The concentration of $S_1$, denoted by $c_1$, is governed by 
\begin{align}
\label{eq:rsb-c1}
    \frac{\partial c_1}{\partial t} =& D \nabla^2 c_1 + k_1^+ - k_1^- c_1 \nonumber \\ 
    & - \sum_{i \in \sigma_1}\left( k_2^+ c_1 c_2^2 - k_2^- c_2^3 \right)\delta(\bx - \bX_i).
\end{align}
The concentration of $S_2$, or $c_2$,  satisfies the equation 
\begin{eqnarray}
\label{eq:rsb-c2}
    \frac{\partial c_2}{\partial t} &=& D \nabla^2 c_2 + k_3^- - k_3^+ c_2 \nonumber \\ 
    && +  \sum_{i \in \sigma_1}\left( k_2^+ c_1 c_2^2 - k_2^- c_2^3 \right)\delta(\bx - \bX_i) \nonumber \\ 
    && - \sum_{j \in \sigma_2} k_4 c_2 \delta(\bx - \bX_j). 
\end{eqnarray}
Because of the linear reactions in the bulk, the chemical system is screened; the nonlinear reactions occur only on the surface of the particles. The inverse screening lengths can be defined as $\lambda_1 = \sqrt{k_1^-/D}$ and $\lambda_2=\sqrt{k_3^+/D}$. Scaling the screening lengths by the radius $a$, we have the non-dimensional screening lengths given by $\Lambda_1 = \lambda_1 a $ and $\Lambda_2 = \lambda_2 a$. For simplicity, we only consider the phoretic motion induced by concentration gradients and set the self-diffusiophoretic coefficients and the rotational diffusiophoretic coefficients to be zero. In the simulations, the number of type 1 particles is the same as type 2,  $N_1=N_2=500$; the area fraction counting both types of particles is $\phi=0.1$.  

In the single and sequential reactions considered in previous sections, the total concentration of all species is conserved, which provides a reference concentration for non-dimensionalization. In the reversible Selkov model, such a simple conservation statement is absent. To non-dimensionalize the system, we define the reference concentration $\tilde{c}=k_1^+a^2/D$, which is obtained by balancing the diffusive term with the bulk production of $S_1$ and taking the radius $a$ of the particles as the length scale. With this, we introduce the following non-dimensional reaction parameters:
\begin{equation}
 \hat{k}_2^\pm = \frac{k_2^\pm \tilde{c}^2}{D} = \frac{k_2^\pm \left(k_1^+\right)^2a^4}{D^3},\quad \hat{k}_3^- = \frac{k_3^-}{k_1^+}, \quad\text{and} \quad\hat{k}_4 = \frac{k_4}{D}.
\end{equation}
The non-dimensional  phoretic coefficients are given by $\hat{\mu}_{ij} = \mu_{ij}\tilde{c}/D$ for $i,j=1,2$.

\begin{figure*}
 \centering
 \includegraphics[width=\textwidth]{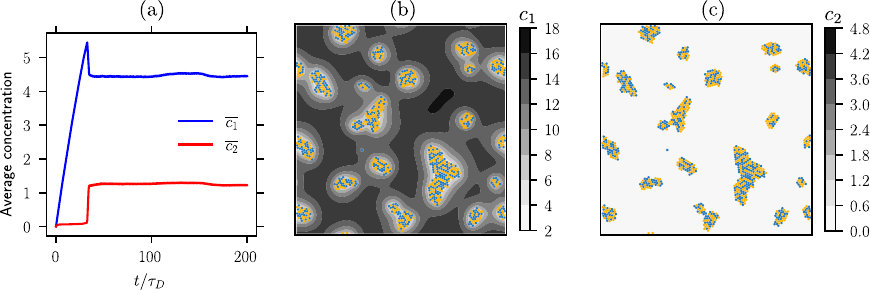}
 \caption{Heterogeneous clusters of the binary Selkov model. (a): The spatially-averaged chemical concentrations vs. the scaled time $t/\tau_D$. (b): The system snapshot at $t/\tau_D=200$ and the background contour plot  denotes the concentration $c_1$. (c): The system snapshot at $t/\tau_D=200$ (same as (b)) and the background contour plot denotes the concentration $c_2$. In this simulation, the self-diffusiophoretic coefficients and the angular diffusiophoretic coefficients are zero. The active velocities of the particles result from the translational diffusiophoretic motion: $\hat{\mu}_{11} = \hat{\mu}_{12}=\hat{\mu}_{21}=-2$ and $\hat{\mu}_{22}=2$. Other physical parameters used are as follows: $\Lambda_1 = 0.1, \Lambda_2=1, \hat{k}_2^\pm =0.04, \hat{k}_3^- = 0.45$ and $\hat{k}_4=4$. Notice that the dimensional screening length of $c_1$ is $10a$, or five particle diameters, while the screening length of $c_2$ is equal to the radius.
}
 \label{fig:rsb}
\end{figure*}

In Fig. \ref{fig:rsb}, we show the  clustering behavior of the reversible Selkov model when $\hat{\mu}_{11} = \hat{\mu}_{12}=\hat{\mu}_{21}=-10$ and $\hat{\mu}_{22}=10$. With these phoretic coefficients, the binary mixture forms localized clusters that contain both types of particles. Since $\hat{\mu}_{11} <0$ and $\hat{\mu}_{21}<0$, particles are attracted to the the sinks of $S_1$, and at long times the clusters are localized in regions of low concentration of $S_1$. The concentration of $S_1$, or $c_1$, is shown in the contour plot in Fig. \ref{fig:rsb}(b). The same clusters are shown in Fig. \ref{fig:rsb}(c) with the contour representing the concentration of $S_2$ ($c_2$). We notice that the concentration gradient of $S_2$ is not visible in  Fig. \ref{fig:rsb}(c) because the dimensional screening length of $c_2$ is equal to the radius of the particle. As a result, gradients of $c_2$ are highly localized at each particle and large scale concentration gradients of $c_2$ can not be established. Within a cluster, type 1 particles produce $S_2$ and subsequently $S_2$ is consumed by type 2 particles.  Because the screening length of $c_1$ is five particle diameters, the gradients of $c_1$ are visible. In the results shown in Fig. \ref{fig:rsb}, the dominant attraction mechanism  is from $\bnabla c_1$, which dictates the clustering dynamics. 

 In the simulation of Fig. \ref{fig:rsb}, the initial concentrations for both $S_1$ and $S_2$ are zero. The spatially averaged concentrations are plotted in Fig. \ref{fig:rsb}(a) as a function of the scaled time. Initially, the bulk production of $S_1$ and $S_2$ dominates and nonzero concentrations of $S_1$ and $S_2$ are gradually established. Once type 1 particles `detect' a finite concentration of $S_1$ and $S_2$, the autocatalytic reaction will be activated and we observe a sharp decline in $\overline{c}_1$ and an increase in $\overline{c}_2$. During this time, the system is also starting to form small clusters. At long times, we observe  mostly steady average concentrations for both $S_1$ and $S_2$ with fluctuations. The small clusters are attracted to each other and at long times form larger clusters as shown in Figs. \ref{fig:rsb}(b) and \ref{fig:rsb}(c).

\begin{figure}
\centering
  \includegraphics[width=2in]{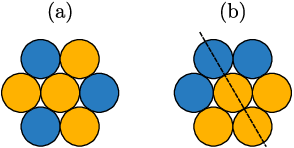}
  \caption{Schematic of two colloidal molecules composed of three particles of type 1 and four particles of type 2. The molecule in (a) has rotational symmetry; the molecule in (b) has a line of symmetry as denoted by the dashed line. }
  \label{fig:molecules_schematic}
\end{figure}

\begin{figure}
\centering
  \includegraphics[width=3in]{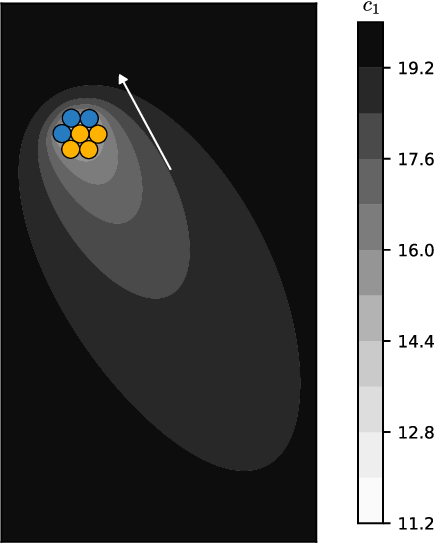}
  \caption{Snapshot of the translating molecule and the concentration field $c_1$. The direction of motion is denoted by the arrow. }
  \label{fig:molecule_translate}
\end{figure}

The binary mixture with Selkov kinetics can also form self-assembled colloidal molecules. In the absence of Brownian motion, colloidal molecules that are stationary or exhibit net motion can be obtained. Using a simple model in which the particles act as sources or sinks for a single otherwise diffusing chemical field, previous studies have shown that stable colloidal molecules of various compositions can be formed \cite{Soto2014,Soto2015}. In contrast, in our current model we have two chemical species and nonlinear reaction kinetics.  For simplicity, we consider in isolation a collection of seven particles in which three are of type 1 (blue in Fig. \ref{fig:selkov-schematic}) and the other four are of type 2 (gold in Fig. \ref{fig:selkov-schematic}). In Fig. \ref{fig:molecules_schematic}, we show examples of two  colloidal molecules that can be formed by taking the same parameters as those in Fig. \ref{fig:rsb}. The molecule in Fig. \ref{fig:molecules_schematic}(a) has rotational symmetry; in the absence of Brownian fluctuations, it can maintain a stationary state. The molecule in Fig. \ref{fig:molecules_schematic}(b) has a line of symmetry and exhibits net translational motion along this line. The blue particles consume $S_1$ and create a concentration gradient with lower concentrations of $S_1$ near the blue particles and higher concentrations near the gold ones (see Fig. \ref{fig:molecule_translate}). Because the gold particles are attracted to regions of low concentrations of $S_1$, the molecule translates in a direction with the blue-particle side at its head (see ESI, Movie 1).

By treating the colloidal particles as a spatially homogeneous system, we note that the chemical system is at a stable fixed point. Denoting the spatially constant densities of type 1 and 2 particles as $n_1$ and $n_2$, respectively, one may write $\partial_t c_1= D\nabla^2 c_1+ k_1^+ - k_1^- c_1 - (k_2^+ c_1c_2^2 - k_2^-c_2^3)n_1$, and $\partial_t c_2=D\nabla^2 c_2+ k_3^- - k_3^+ c_2 + (k_2^+ c_1c_2^2 - k_2^-c_2^3)n_1 - k_4 c_2 n_2$. At steady state for system parameters of Fig. \ref{fig:rsb},  these equations give $c_1 \approx 19.764$ and $c_2 \approx 0.087$. A linear stability analysis about this fixed point reveals that the chemical system is stable to inhomogeneous concentration perturbations for all wavenumbers, which suggests that the instability (i.e., emergence of a non-homogeneous density state) shown in Fig. \ref{fig:rsb} results from the clustering instabilities induced by phoretic interactions.  In a full description that includes the equation of motion for the inhomogeneous density fields, for appropriate system parameters, the chemical system itself may be unstable and the interaction of the chemical instability with the clustering instability may give rise to more complex bifurcations and dynamics.  

The above results provide some simple examples of the phenomena that can be seen in systems where the chemical mechanism involves nonlinear kinetics on specific colloidal particles, along with chemical reactions on other colloids and in the surrounding fluid. The possible types of behavior is far richer than that explored here. For example, the formation of clusters with specific compositions and surface reaction kinetics constitute a compartmentalized reaction-diffusion system where portions of the reaction network reside in spatially localized compartments that communicate by diffusion. Depending on the diffusion length relative to the distance between compartments, the system dynamics can change its bifurcation structure.~\cite{CK2000} A laboratory example is the Belousov-Zhabotinsky reaction carried out in miroemulsions.~\cite{VE2001} In this system different parts of the mechanism are confined to the different phases in the emulsion, and the dynamics depends strongly on the composition of the microemulsion. In biological systems reactions are often compartmentalized in specialized organelles and communicate with reactions in the surrounding cytoplasm through diffusion~\cite{GK1973}. Dynamic clustering in active colloidal systems, where the reaction network distributed over different system constituents, is expected to display a similarly rich bifurcation structure that now depends on the dynamics of the clustering process. While such phenomena have not been explored in this paper, they deserve further study.

\section{Conclusion}
\label{sec:conclusion}
In this paper, we have considered the collective motion and self-organization of chemically active colloids propelled by complex chemical reactions involving multiple chemical species and potentially nonlinear kinetics. The multiple chemical species are coupled by a network of chemical reactions.  In general, the results have shown that active colloids can form clusters in the presence of phoretic attractions. The characteristics of the clusters, however, depend on the reaction kinetics and the resulting chemical fields and phoretic coefficients. In binary mixtures, we have shown that diverse clustering behaviors can be obtained including the clustering of one type of colloid while the other type is dispersed, heterogeneous clusters containing both types of particles, and  heterogeneous clusters with one type of colloid on the periphery. We have also shown that chemical screening typically leads to localized clusters in the dilute regime. With a single chemical species, the tunable parameters for self-organization are limited. By allowing multiple chemical reactions both on the colloids and in the surrounding fluid, our study shows that  the design space of colloidal self-organization can be broadened by including chemical reaction kinetics as an additional parameter.

The complex chemical reaction kinetics considered in our work can be implemented using enzyme-coated colloidal particles or natural cells. In the context of designing micro- and nano-motors for biomedical applications, enzymatic reactions that make use of readily available biofuel is desirable. For example, an endogenous enzyme-powered Janus platelet micro-motor is shown to exhibit phoretic motion due to the uneven decomposition of urea in biofluids\cite{tang2020enzyme}. Micro- and nano-motors powered by other enzymatic reactions have also been realized\cite{gao2013multi,Sch2017,Dey2015,Sch2015,Ma2016,Ma2016b,Zhao2018,toebes2019spatial,arque2019intrinsic,tang2020enzyme}. The enzymatic reactions that can be used for self-propulsion and self-organization may depend on the biological environment and whether the required biofuel species are present in such an environment. As a result, it is beneficial to develop a variety of enzyme-powered motors that make use of different reaction schemes.

While we focused on demonstrating the diverse clustering behaviors that can be achieved, our more general model outlined in section \ref{sec:model} includes nonlinear kinetics and bulk reactions. In both the sequential and nonlinear models considered in this paper, the bulk reaction kinetics are linear. We note that generic nonlinear reaction kinetics in the bulk can be employed to establish external gradients that influence the clustering dynamics of the particles. In the case of linear bulk kinetics, the resulting multiple screening lengths can also be used to control the system dynamics. In addition to phenomena related to compartmentalized reaction-diffusion systems discussed above, the research reported here can also form the basis for of chemical computation in active colloidal systems. The surface reactions (especially those that are nonlinear) can be constructed to be logic gates~\cite{CHK2023}, and clusters of colloids supporting different gates can lead to the self assembly of logic circuits. In this way it may be possible for the system to carry out chemical computations to influence its dynamics to perform specific tasks.

To study the self-organization of active colloids mediated by complex chemical reactions, we presented a minimal computational model.   In constructing our model, for simplicity we used a monopole approximation for the chemical fields following previous studies and we have neglected hydrodynamic interactions. To account for the near-field chemical interactions, one needs to solve the reaction-diffusion equations subject to surface reaction boundary conditions. For constant or linear surface reaction kinetics in the absence of bulk reactions, the reaction-diffusion equation of a single chemical species in the presence of one or two active particles have been considered\cite{michelin2013spontaneous,michelin2014phoretic,michelin2015autophoretic,michelin2017geometric,nasouri2020exact,nasouri2020exactJFM}. Mesoscopic particle-based~\cite{K08} and other numerical methods~\cite{varma2019modeling,rojas2021hydrochemical} can be used to take into account full complex chemical dynamics as well as hydrodynamic interactions.

\section*{Author Contributions}
Z. P. and R. K. contributed at all stages of this work. Z. P. developed the simulation software and performed simulations.

\section*{Conflicts of interest}
There are no conflicts to declare.

\section*{Acknowledgements}
This work was supported in part by the Natural Sciences and Engineering Research Council (NSERC) of Canada. Computations were performed on SciNet HPC Consortium computers. SciNet is funded by the Canada Foundation for Innovation, the Government of Ontario, the Ontario Research Excellence Fund, and the University of Toronto.

\appendix
\section{Numerical and Brownian dynamics simulations}
Consider a collection of $N$ spherical particles  in  a square simulation box $[-L, L] \times [-L, L]$ with periodic boundary conditions in both directions. To evolve the reaction-diffusion equations numerically, we discretize space using a uniform grid of spacing $h=2L/M$, where $M$ is the number of intervals in each direction. The reaction-diffusion equations are discretized using second-order centered finite differences. Taking the second equation in the sequential reaction scheme as an example, the discretization gives 
\begin{eqnarray}
       &&\frac{\od  c_2[i,j]}{\od t} = D  \frac{c_2[i+1,j] - 2 c_2[i,j] + c_2[i-1,j]}{h^2}\nonumber \\ 
       &&+ D  \frac{c_2[i,j+1] - 2 c_2[i,j] + c_2[i,j-1]}{h^2} \nonumber \\ 
       && - k_{b,1} c_2[i,j] + k_{b,2} c_3[i,j]\nonumber \\ 
       && + \kappa_1 \sum_{k=1}^{N_1} \frac{c_1[i_k, j_k]}{h^2} - \kappa_2 \sum_{k=N_1+1}^{N} \frac{c_2[i_k, j_k]}{h^2},
\end{eqnarray}
where $c_1$ is now the values at grid points and $c_1[i,j]$ denotes the node value at indices ($i,j$), and $i_k$ and $j_k$ denotes the index of the grid cell in which a particle resides. Because we need to know the particle locations in order to evolve the reaction-diffusion equation in time, an occupancy matrix (bit array) is maintained for each particle type. We then discretize time using the first-order explicit Euler scheme. At each time step, the reaction-diffusion equations are first evolved, from which we calculate the active linear and angular velocities. With these velocities, we then perform Brownian dynamics simulations that will  update the particle positions and orientations. 

We simulate the Langevin equations of motion using Brownian dynamics, where the fluctuating velocities are treated using the Euler-Maruyama scheme.  The hard-sphere interactions among particles are treated using a potential-free geometric optimization approach and all possible collisions during a discrete time step are resolved simultaneously. This scheme requires solving a linear complementarity problem, where a projected gradient descent method is used (see Refs.\cite{Yan2019,Yan2020} for details). The simulations are performed using an in-house program that runs on CUDA-enabled NVIDIA GPUs.





\bibliography{rsc} 
\bibliographystyle{rsc} 





\end{document}